\documentclass[12pt]{article}
\usepackage[english]{babel}
\usepackage{amsmath} 
\usepackage{amssymb}
\usepackage{amsfonts}
\usepackage{amsthm}
\usepackage{multicol}
\usepackage{pdflscape}
\usepackage{multirow}
\usepackage[table,xcdraw]{xcolor}
\usepackage{array}
\usepackage{xcolor}
\usepackage{ragged2e}
\usepackage{tikz}
\usepackage{graphicx}
\usepackage{gensymb}

\newtheorem{remark}{Remark}

\usepackage{hyperref}
\usepackage{url}
\usepackage{enumerate}
\usepackage{graphicx}
\usepackage{epstopdf} 
\usepackage{geometry}
\usepackage{cite}
\usetikzlibrary{shapes, shadows, arrows}
\tikzstyle{block}=[draw,rectangle,fill=blue!5,text width=12 em,text centered, minimum height=12mm, node distance=5 em]
\tikzstyle{line} = [draw,-latex']
\begin{document}
	\title{\textbf{Understanding Hepatitis B Virus Infection through Hepatocyte Proliferation and Capsid Recycling}}
	\date{} 
	\author{Rupchand Sutradhar and D C Dalal}
	\maketitle
\section{Abstract}
Proliferation of uninfected as well as infected hepatocytes and recycling of DNA-containing capsids  are two major mechanisms playing significant roles in the clearance of hepatitis B virus (HBV) infection. In this study, the temporal dynamics of this infection is investigated through two in silico bio-mathematical models considering both proliferation of hepatocytes and recycling of capsids. Both models are formulated on the basis of a key finding in existing literature: Mitosis of infected yields in two uninfected progenies. In the first model, we examine regular proliferation (occur continuously), while the second model deals with the irregular proliferation (happen when the total number of  liver cells  decreases to less than 70\% of its initial volume). The models are calibrated with the experimental data obtained from an adult chimpanzee. Results of this study suggest that when both hepatocytes proliferate with equal  rate, proliferation aids the individual in a rapid recovery from the acute infection whereas in case of chronic infection, the severity of the infection increase if the proliferation occur frequently. On the other hand, if the infected cells proliferate at a slower rate than uninfected cells, the proliferation of uninfected hepatocytes contributes to  increase  the infection, but the proliferation of infected hepatocytes acts to reduce the infection from the long-term perspective. Furthermore, it is also observed that the differences between the outcomes of regular and irregular proliferations are substantial and noteworthy.
	\vspace{0.5cm}\\
Keywords:  Hepatitis B, capsids, recycling, chronic, cellular proliferation, irregular proliferation.
\section{Introduction}
Hepatitis B virus (HBV) infection stands as an extensively widespread infection  with more than 300 million chronically infected individuals  across the world. This viral infection  leads to approximately 1 million fatalities annually despite the fact that it is preventable and treatable \cite{Hepb_2021}. HBV is a partially double-stranded, enveloped  DNA virus. This viruses mainly target hepatocytes, the primary cells of the liver. Even with the availability of antiviral drugs (interferons (IFN)-alpha-2a, pegylated (PEG)-IFNalpha-2a (immune system modulators), and some nucleoside analogues, such as lamivudine, adefovir, entecavir, telbivudine, and tenofovir) and  effective vaccine, HBV remains a significant public health concern. However, the long-term use of these medications can lead to the development of drug resistance, diminishing their effectiveness over time \cite{2011_zoulim_hepatitis}. The amalgamation of in vitro experiments and in vivo explorations in ducks, woodchucks, mice, and chimpanzees has greatly advanced our knowledge of HBV infection and its interplay with the immune system. Extensive efforts have been dedicated over the past three decades to gain a comprehensive understanding of HBV\cite{1996_Nowak,2002_lewin_hepatitis,2003_Wodarz,2006_Murray,2006_Guidotti,2007_Ciupe,2008_Min,2009_Eikenberry,2010_Hews,2011_Jun_nakabayashi,2014_Chen,2018_fatehi_nkcell,2021_liu_age,2022_Tu_mitosis,2021_prifti,2018_ko_hepatitis}. Especially, regarding both acute and chronic HBV infection in humans, our knowledge remains still restricted due to the paucity of available data. Even so, there exists some data \cite{2009_asabe_size,PPR:PPR672105}, and using this data with mathematical models, possible mechanisms underlying HBV infection are investigated \cite{2023_sutradhar_intracellular,2023_Sutradhar_fractional}. 

Cellular proliferation or mitosis cell division stands as one of the  pivotal biological process influenced by  this viral infection as HBV promotes cell proliferation through HBx-induced microRNA-21 in hepatocellular carcinoma (HCC) \cite{2014_damania_hepatitis}.  In the literature, a limited number of studies are dedicated in investigating the proliferation of hepatocytes. For example, in the year  2010, Hews et al. \cite{2010_Hews} introduced a HBV infection dynamics model considering the logistic growth of uninfected hepatocytes and standard incidence function. However, it's worth noting that their model incorporated numerous assumptions while also omitting several crucial pieces of information that could have been included in their model. Taking into account the supposition that infected hepatocytes also undergo proliferation at a rate comparable to or lower than that of healthy hepatocytes during the infection, Hews et al. \cite{2021_hews_global} modified their previously proposed model  \cite{2010_Hews} and obtained some new outcomes. Dahari et al. \cite{2009_dahari_modeling} suggested that variations in proliferation between uninfected and infected hepatocytes could offer  a rich explanation for the diverse patterns that are observed in the decline of viral loads. Some studies have also addressed the concept of proliferation of infected hepatocytes (POIH)  with the assumption that POIH only can produce the infected hepatocytes \cite{2017_carracedo_understanding,2009_reluga_analysis}. Based on the data collected regarding cellular proliferation, it is noted that  three  different scenarios may occur when an infected hepatocyte undergoes proliferation (mitosis). 
\begin{enumerate}
	\item Two uninfected daughter cells.
	\item One uninfected and one infected daughter cell.
	\item Two infected daughter cells.
\end{enumerate}
\begin{figure}[h]
	\centering
	\includegraphics[height=8cm,width=16cm]{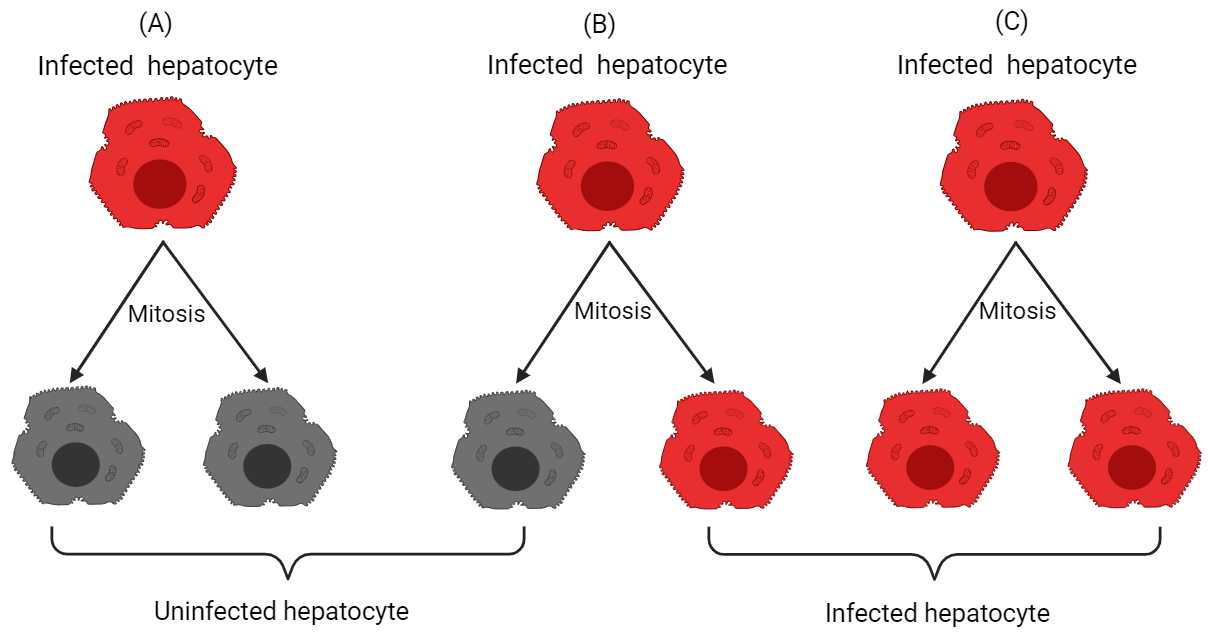}
	\caption{Proliferation of an infected hepatocyte results (A): Two uninfected hepatocytes, (B): One uninfected and one infected hepatocytes, (C): Two infected daughter cells. This illustration is created with BioRender.com.}
	\label{liver failure}
\end{figure}
The fate of viral DNA after cell mitosis is a matter of controversy. The roles of proliferation in the clearance of both acute and chronic HBV infections continue to be subjects of ongoing debate \cite{2016_seeger_hbv,2018_allweiss_proliferation,2010_lutgehetmann_vivo}. 
During mitosis cell division, when the nuclear membrane is reformed, there is a possibility of cccDNA loss if the cccDNAs fail to reintegrate into the nucleus.
Also, if the count of cccDNA within infected hepatocytes is extremely low, the progeny will be devoid of cccDNA, leading to the resolution of acute HBV infection \cite{2017_goyal_role}. Thus, the involvement of this additional mechanism is crucial in achieving the clearance of acute infection. Recently, Tu et al. \cite{2022_Tu_mitosis} corroborate that the mitosis cell division of hepatitis B virus-infected cells leads to the generation of two uninfected daughter cells. This research team also suggests that mitosis of infected hepatocytes offers a potent avenue for addressing HBV persistence. Murray et al. \cite{2015_murray_silico} also highlighted the significance of cccDNA loss during cellular proliferation in the context of non-destructive clearance of HBV infection. Using experimental data from six HBV-infected patients, Goyal et al. \cite{2017_goyal_role} further affirmed that the resolution of acute HBV infection is probably contingent on the cellular proliferation of an infected cell leading to the generation of two uninfected daughter cells. To the best of our knowledge, no  relevant article has been found in the existing literature that addresses the impacts of POIH on chronic infection.
In this study, we mainly focus on the proliferation of uninfected as well as infected hepatocytes along the important mechanism: recycling of HBV DNA-containing capsids which is recently introduced by Sutradhar and Dalal \cite{2023}. 
Based on the conclusion of the previous works \cite{2017_goyal_role,2022_Tu_mitosis}, we proceed with the fact that POIH produces  two healthy progeny.  In this study, the following things will be discussed:
\begin{itemize}
	\item  When both uninfected and infected hepatocytes proliferate with equal rate.
	\begin{itemize}
		\item Effects of  proliferation rates on the acute infection.
		\item Effects of  proliferation rates on the chronic infection.
	\end{itemize}
	\item When both uninfected and infected hepatocytes proliferate with different rates.
 	\begin{itemize}
		\item Effects of  proliferation rates on the acute infection.
		\item Effects of  proliferation rates on the chronic infection.
	\end{itemize}
	\item Proliferation is not a continuous
	process: regular proliferation vs irregular proliferation.
\end{itemize}
Considering all the aforementioned factors, we put forth a mathematical model that incorporates both the proliferation of infected hepatocytes (produce two uninfected daughter cells) and recycling of capsids. Due to high non-linearity, the proposed model is solved through numerical methods, and the outcomes for various scenarios are comprehensively elucidated. 

\section{Model formulation and analysis} \label{Sec: Model formulation and analysis}
\begin{figure}
	\centering
	\includegraphics[height=8cm, width=12cm]{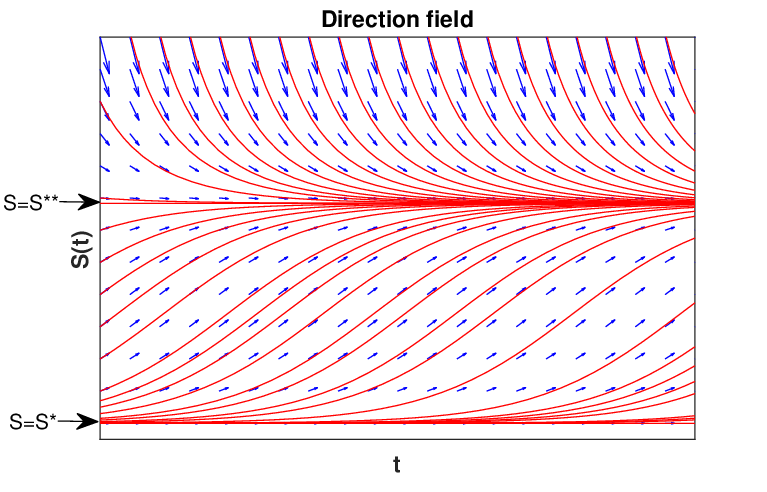}
	\caption{Direction field of inequation \eqref{Eqn: of S}. }
	\label{direction field}
\end{figure}

In order to analyze the HBV infection focusing  on  the role of proliferation of susceptible and infected hepatocytes to the HBV infection, a new dynamics model is proposed in this study. The total number of liver cells of a healthy adult is denoted by $T$ which contains both hepatocytes and nonparenchymal cells (cells in the liver that cannot be infected). It is estimated that for an adult the total number of liver cells $(T)$ is  $2 \times 10^{11}$ and out of them only $60\%$ are virus-targeted liver cells \cite{2007_Ciupe_Role,2017_goyal_role,2001_kmiec_cooperation}. This type of hepatocytes (virus-targeted cells)  are grouped into two classes in this model: susceptible hepatocytes $(X)$ and infected hepatocytes $(Y)$. On the other hand, as model compartments, we take into account two components of the virus life cycle: (i) rcDNA containing capsids $(D)$, and (ii) complete virions $(V)$. In addition, it is assumed that the susceptible and virus infected hepatocytes  proliferate following the logistic growth law with proliferation rate $r$ and $\rho$, respectively. It is considered that the infection occurs at a constant rate $k$. The parameter $\mu$ represents the natural death rate of susceptible hepatocytes.  Accordingly, the susceptible hepatocytes adhere to the following dynamical equation:
\begin{equation} \label{eq_susceptible hepatocytes}
	\begin{split}
		\dfrac{dX}{dt}&=r X \left(1-\dfrac{X+Y+\theta T}{T}\right)+2 \rho Y \left(1-\dfrac{ X+Y+\theta T}{T}\right)-k X V-\mu X=f_1(X,Y,D,V),
	\end{split}
\end{equation}
where $\theta$ denotes the fraction of nonparenchymal cells. 
The term $r X \left(1-\dfrac{X+Y+\theta T}{T}\right)$ represents the fact that upon proliferation, an uninfected cell always produces  uninfected cells. The second term of RHS of equation  \eqref{eq_susceptible hepatocytes} \textit{i.e.} $2 \rho Y \left(1-\dfrac{ X+Y+\theta T}{T}\right)$ indicates that one infected hepatocyte proliferates and results in two uninfected hepatocytes. We denote the per capita death rate of infected hepatocytes with the parameter $\delta$.  The corresponding governing  equation of infected compartment is given by
	\begin{equation} \label{eq_infected hepatocytes}
	\begin{split}
		\dfrac{dY}{dt}&=k X V-\rho Y \left(1-\dfrac{X+Y+\theta T}{T}\right)-\delta  Y=f_2(X,Y,D,V).
	\end{split}
   \end{equation}
 The model also incorporates the recycling of capsids.
   In equation \eqref{eq_capsids}, the parameter $\gamma$ represents the recycling rate or capsids to capsids production via recycling whereas $\alpha$ indicates the  volume fraction of newly produced  capsids which are responsible for new virus production.
     During the infection, it is considered that the rcDNA-containing capsids are generated from infected hepatocytes with rate $a$. Mathematically, the dynamical change of  capsid compartment is represented in the  following way:
	\begin{equation} \label{eq_capsids}
	\begin{split}
		\dfrac{dD}{dt}&=a Y+\gamma(1-\alpha)D-\alpha\beta D-\delta D=f_3(X,Y,D,V),
	\end{split}
    \end{equation}
where the production rate of virus from capsid is described by $\beta$ and $c$ performs as decay rate of viruses. Here, equation \eqref{eq_virus} describes the reaction equation for virions compartment.
\begin{equation} \label{eq_virus}
	\begin{split}
		\dfrac{dV}{dt}&=\alpha \beta D-cV=f_4(X,Y,D,V). 
	\end{split}
\end{equation}
\noindent
Clearly, the functions $f_i(X,Y,D,V),~i=1,2,3,4$ are polynomial functions of $X,Y,D$ \& $V$. The initial conditions $(X_0,Y_0,D_0,V_0)$ are always non-negative.  Each $f_i$ is continuous and satisfies the well-known Lipschitz condition  for uniqueness of solution. Hence, the initial value problems \eqref{eq_susceptible hepatocytes}, \eqref{eq_infected hepatocytes}, \eqref{eq_capsids}, \eqref{eq_virus} has unique solution in any interval $[0, b), b\in \mathbb{R}^+$. With non-negative initial conditions, the solutions of the proposed model will remain non-negative. In order to show the boundedness of the solution with respect to non-negative initial conditions, we assume that 
proliferation rate of uninfected hepatocytes ($r$) is greater than the proliferation rate of infected hepatocytes ($\rho$) \textit{i.e.} $r\geq \rho$. 
	Consider a new variable $S=X+Y$ then $\dfrac{dS}{dt}=r X \left(1-\dfrac{X+Y+\theta T}{T}\right)+ \rho Y \left(1-\dfrac{ X+Y+\theta T}{T}\right)-\mu X-\delta Y.\\$
	Using the above assumption, we have\\
	\begin{align} \label{Eqn: of S}
             \dfrac{dS}{dt}\leq rS\left(1-\dfrac{S+\theta T}{T}\right)-\mu S.
	\end{align}
	The equilibrium points of inequation \eqref{Eqn: of S} are $S^*=0$ and  $S^{**}= T \left(\dfrac{r(1-\theta)-\mu}{r}\right).$ In order to show boundedness of inequation \eqref{Eqn: of S}, the following cases are considered:
	\begin{itemize}
		\item \textbf{Case 1:} Let  $S_0$ be the initial condition satisfying the relation $S^*\leq S_0\leq S^{**}$ which implies $0<S_0<T \left(\dfrac{r(1-\theta)-\mu}{r}\right)$. Upon simplification, we see that 
		\begin{align}\label{Equ: S0 only}
			rS_0\left(1-\dfrac{S_0+\theta T}{T}\right)-\mu S_0\geq 0.
		\end{align}  
		Combining, inequation \eqref{Eqn: of S} and \eqref{Equ: S0 only}, one can observe that
		\begin{align} \label{Eqn: of S0}
			\dfrac{dS}{dt}\leq rS_0\left(1-\dfrac{S_0+\theta T}{T}\right)-\mu S_0\geq 0.
		\end{align}
	
 \begin{itemize}
 	\item \textbf{Subcase-1: } If $\dfrac{dS}{dt}\leq 0$, then the solution converges to $S^*$.
 	\item \textbf{Subcase-2: } If $0<\dfrac{dS}{dt}\leq rS_0\left(1-\dfrac{S_0+\theta T}{T}\right)-\mu S_0$, then the solution $S(t)$ converges to $S^{**}$.
 \end{itemize}
 \item \textbf{Case-2:} When $S_0>S^{**}$, then 
 $$ rS_0\left(1-\dfrac{S_0+\theta T}{T}\right)-\mu S_0 < 0,$$ which indicates that $\dfrac{dS}{dt}<0$. Therefore, the solution $S(t)$ decreases and converges to the equilibrium point $S^{**}$.
\end{itemize}
Hence, in all cases $S(t)$ is bounded which implies that $X(t)~\text{and}~Y(t)$ both are bounded. The associated direction field of the inequation \eqref{Eqn: of S} is given by Figure \ref{direction field}.  Subsequently, when $R_s=\alpha  \beta -(1-\alpha) \gamma + \delta>0$, by employing the same approach outlined in the article of  Sutradhar and Dalal \cite{2023_sutradhar_recycling}, one can easily establish the boundedness of both $D(t)$ and $V(t)$.

Based on the proliferation rate of healthy and unhealthy hepatocytes, two different scenarios are analyzed here. In the first case, it is assumed that the both type of liver cells proliferate with the same rate \textit{i.e.} $r=\rho$.  In the second, we consider that as compared to healthy hepatocytes, infected hepatocytes proliferate at a lower rate. In both cases, we consider non-negative initial conditions. Since  in both cases, the  covalently closed circular DNA (cccDNA) pool contained in the  infected hepatocytes is disappeared after mitosis,  this type of model is sometime refereed to as the ``cccDNA loss model'' \cite{2022_Tu_mitosis}.

\section{Parameter estimation and initial condition} 
The parameter values are taken from the existing literature and shown in Table \eqref{Table-Parameter values}. The quantity $(1-\theta)$ denotes the fraction of the total liver cells that are hepatocytes and primary target of the viruses.  Based on the prior studies \cite{2008_kuntz_systemic,2002_sherlock_chapter}, it is considered that  $\theta = 0.4$ due to the fact that  60\% of liver cells are  hepatocytes. The value of proliferation rate of infected hepatocytes $(\rho)$ is constrained to be between 0.001 and 0.35/day \cite{2017_goyal_role}.  The value of proliferation rate $(r)$ of uninfected hepatocytes lies between 0.001 and 3.4/day \cite{2009_reluga_analysis}.


\begin{table}[h!]
	\caption{ \label{Table-Parameter values} Estimation of parameters.}
	\begin{center}
		\footnotesize
		\begin{tabular}{c |l| c| c| c}
			\hline
			\cellcolor{blue!10}Parameters  & ~~~~~~~~~~~~~~\cellcolor{blue!10}Descriptions  	&\cellcolor{blue!10} Values 		&\cellcolor{blue!10} Units 	 & \cellcolor{blue!10}Sources  \\ [.5ex]
			\hline
			\cellcolor{red!20}$r$ & \cellcolor{red!20}Proliferation rate of uninfected hepatocytes & \cellcolor{red!20}$0.001-3.4$ &\cellcolor{red!20} $\mbox{day}^{-1}$ &\cellcolor{red!20}\cite{2017_goyal_role}\\
			\hline
		\cellcolor{blue!10}	$\rho$ &\cellcolor{blue!10}Proliferation rate of infected hepatocytes&\cellcolor{blue!10} $0.001-0.35$ &\cellcolor{blue!10} $\mbox{day}^{-1}$ &\cellcolor{blue!10}\cite{2009_reluga_analysis,2009_dahari_mathematical}\\
			\hline
				\cellcolor{red!20}$k$&	\cellcolor{red!20}Infection rate &	\cellcolor{red!20} $0.55\time10^{-10}$ &	\cellcolor{red!20}$\text{mL}~ \text{copies}^{-1}\text{day}^{-1}$&	\cellcolor{red!20} \cite{2017_goyal_role}\\
			\hline
			\cellcolor{blue!10}$T$& \cellcolor{blue!10}Total number of liver cells & \cellcolor{blue!10} $2\times10^{11}$ &\cellcolor{blue!10} &\cellcolor{blue!10}\cite{2021_hews_global,2017_ogoke_bioengineering}\\
			\hline
			\cellcolor{red!20}	$\theta$&	\cellcolor{red!20}Fraction of liver cells that cannot be infected &	\cellcolor{red!20}0.4 &	\cellcolor{red!20} unitless &	\cellcolor{red!20}\cite{2017_goyal_role} \\
			\hline
		\cellcolor{blue!10}	$\mu$& \cellcolor{blue!10}Natural death rate of uninfected hepatocytes & \cellcolor{blue!10} $0.01$      & \cellcolor{blue!10}$\mbox{day}^{-1}$   &\cellcolor{blue!10} \cite{2007_Dahari}\\
			\hline
				\cellcolor{red!20}$a$			& 	\cellcolor{red!20}Production rate of capsid  	& 	\cellcolor{red!20}150     &	\cellcolor{red!20}$\mbox{capsids  cell}^{-1}\mbox{day}^{-1}$    &	\cellcolor{red!20}\cite{2006_Murray}		\\
			\hline
		\cellcolor{blue!10}	$\beta$ 	&  \cellcolor{blue!10}Production rate of virus from capsid 				& \cellcolor{blue!10}0.87    &\cellcolor{blue!10}  $\mbox{day}^{-1}$   &\cellcolor{blue!10}\cite{2006_Murray}				\\
			\hline
			\cellcolor{red!20}	$\delta$	&	\cellcolor{red!20}Death rate of infected hepatocytes and capsids   &	\cellcolor{red!20} 0.053     &	\cellcolor{red!20}  $\mbox{day}^{-1}$    &	\cellcolor{red!20}\cite{2006_Murray}\\
			\cellcolor{red!20}	& 	\cellcolor{red!20}hepatocyte \& capsid &	\cellcolor{red!20}      & 	\cellcolor{red!20}     &	\cellcolor{red!20}					\\ 
			\hline
		\cellcolor{blue!10}	$c$ 		&\cellcolor{blue!10}Death rate of virus						& \cellcolor{blue!10}3.8     & \cellcolor{blue!10} $\mbox{day}^{-1}$   &\cellcolor{blue!10}\cite{2006_Murray}						\\
			\hline
			\cellcolor{red!20}	$\alpha$ 	&	\cellcolor{red!20}Volume fraction of HBV capsids 			&	\cellcolor{red!20}$0\leq \alpha\leq 1$      &	\cellcolor{red!20} unitless    & 	\cellcolor{red!20}--			\\
			\hline
			\cellcolor{blue!10}$\gamma$ 	&  \cellcolor{blue!10}Capsid to capsid production rate or 			& \cellcolor{blue!10} 0.6931     &\cellcolor{blue!10}$\mbox{day}^{-1}$   & \cellcolor{blue!10}\cite{2015_Murray}			\\
		\cellcolor{blue!10}	& \cellcolor{blue!10}recycling rate &\cellcolor{blue!10} &\cellcolor{blue!10} &\cellcolor{blue!10}\\
			\hline
		\end{tabular}
	\end{center}
\end{table}
\section{Model calibration}
Once the proposed model is validated with experimental data, it becomes a powerful tool for analyzing the infection dynamics under various conditions. In order to validate our proposed model, the experimental data of HBV DNA from a male chimpanzee  are utilized.  This chimpanzee was labeled as A2A007 and was inoculated with $10^4$ GE of HBV. The age and the body weight of this chimpanzee were 5.1 years and  17.7 kg, respectively.  These data have been previously published in the work of Asabe et al. \cite{2009_asabe_size}. In the study carried out by this group, the experiment was done on a cohort of nine healthy adult  chimpanzees. Ethical guidelines governing the use and care of the chimpanzees were meticulously followed during the study. 

\begin{figure}[h!]
	\centering
	\includegraphics[height=9cm,width=15cm]{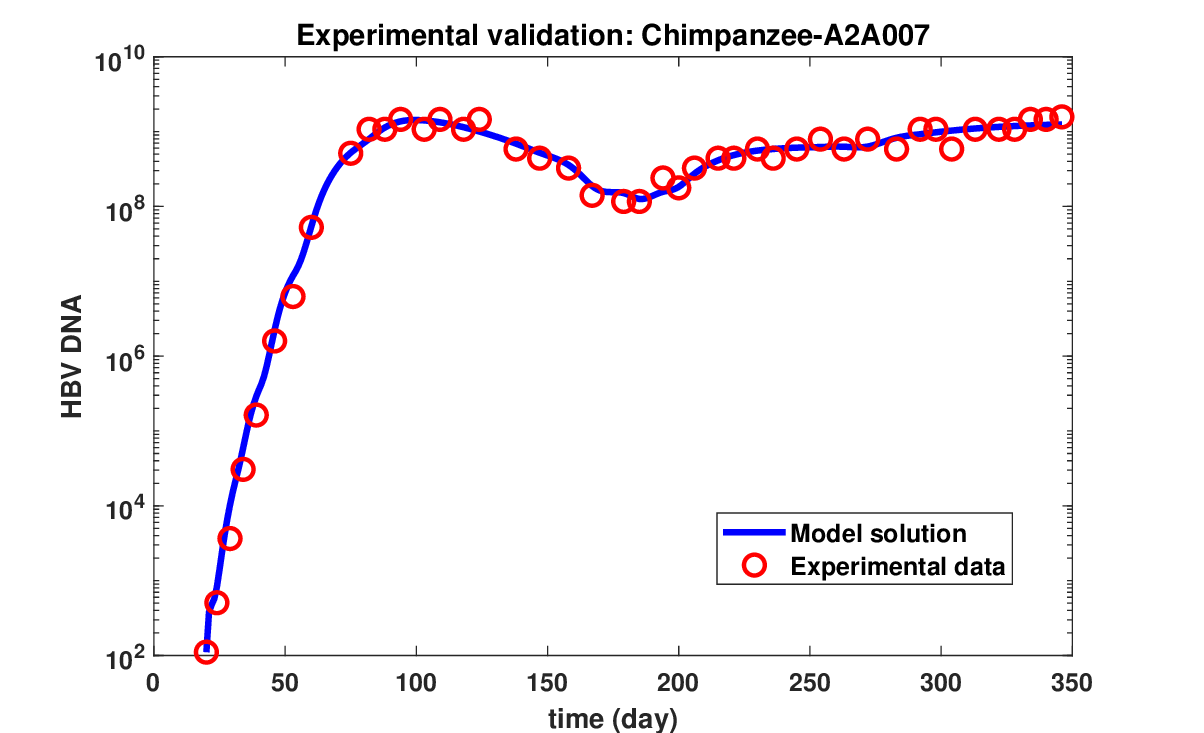}
	\caption{Comparison of model solution  with  experimental data. Solid blue line represents model solution whereas red circles reflect the experiment data.}
	\label{experimental_data}
\end{figure}

Analyzing the experimental data of Chimpanzee A2A007, significant variations in HBV DNA levels were noted precisely at time intervals $t=82$ and $t=185$ days. To delineate and characterize these three sudden changes, we need to divide the time course into three distinct sub-intervals, namely $[20, 82]$, $[82, 185]$, and $[185, 346]$. The model parameters are estimated in each sub-interval minimizing the sum square error (SSE) objective function. The SSE function is defined as
\begin{align}
	SSE=\sum_{i=1}^{N}\left[Q_i-q(i)\right]^2,~~ N=\mbox{total number of experimental data,} 
\end{align}  
where model solution and experimental data are denoted by $Q_i$ and $q(i)$,  respectively. This optimization process is carried out through the utilization of the 'fminsearchbnd' built-in MATLAB function.
 The upper and lower bound of the parameters are collected from the existing  literature. The estimated values of each parameter in each time interval are shown in Table \ref{Estimated value of paramters}.
The  solution of our model and the experimental data are compared in Figure \ref{experimental_data}, and it is  observed that the model solution exhibits a robust agreement with the experimental data.

\begin{table}[]
	\begin{center}
	\begin{tabular}{|c|ccc|}
		\hline
		\multirow{2}{*}{\begin{tabular}[c]{@{}c@{}}Estimated \\ Parameters\\ \end{tabular}} & \multicolumn{3}{c|}{Time interval}                                       \\ \cline{2-4} 
		& \multicolumn{1}{c|}{20-82}    & \multicolumn{1}{c|}{82-185}   & 185-346  \\ \hline
		$r$                                                                                      & \multicolumn{1}{c|}{0.100558} & \multicolumn{1}{c|}{0.150707} & 3.374189 \\ \hline
		$\rho$                                                                                   & \multicolumn{1}{c|}{0.010721} & \multicolumn{1}{c|}{0.062152} & 0.540862 \\ \hline
		$k$                                                                                      & \multicolumn{1}{c|}{7.96E-11} & \multicolumn{1}{c|}{5.83E-11} & 3.2E-10  \\ \hline
		$a$                                                                                      & \multicolumn{1}{c|}{108.2001} & \multicolumn{1}{c|}{40.97998} & 10.00146 \\ \hline
		$\gamma$                                                                                 & \multicolumn{1}{c|}{2.758616} & \multicolumn{1}{c|}{2.107031} & 1.845693 \\ \hline
		$\beta$                                                                                  & \multicolumn{1}{c|}{1.766558} & \multicolumn{1}{c|}{0.887594} & 1.283164 \\ \hline
		$\eta$                                                                                   & \multicolumn{1}{c|}{0.764174} & \multicolumn{1}{c|}{0.702477} & 0.640821 \\ \hline
		$\mu$                                                                                    & \multicolumn{1}{c|}{0.027786} & \multicolumn{1}{c|}{0.198765} & 0.120172 \\ \hline
		$\delta$                                                                                 & \multicolumn{1}{c|}{0.06998}  & \multicolumn{1}{c|}{0.050187} & 0.049513 \\ \hline
		$c$                                                                                      & \multicolumn{1}{c|}{0.147755} & \multicolumn{1}{c|}{0.347936} & 0.241115 \\ \hline
	\end{tabular}
\end{center}
\caption{Estimated value of paramters.}
\label{Estimated value of paramters}
\end{table}

\section{Optimal criteria for maintaining smoothly the essential functions of liver } \label{smoothness of the liver}
In literature, many authors have proposed different types  of HBV infection dynamics model without considering the  lower bound of uninfected hepatocytes and upper bound of infected hepatocytes for survival of the patients. During liver transplantation, resections that leave 20–30\% of the original liver volume are considered safe and allow  the donor liver to regenerate \cite{2012_gruttadauria_early,2012_guglielmi_much}.
According to the  experimental study of Kishi et al. \cite{2009_kishi_three} on more than 300 right lobe hepatectomy patients, it is come to  know that the  liver remnant volumes greater than 20\% are sufficient  to resect the liver safely. So, it is very important to maintain a certain level of uninfected hepatocytes throughout the infection period. As per the discussion, it is assumed that a minimum  20\% of the total hepatocytes (excluding the population of non-parenchymal cells) or 52\% with respect to entire liver size is required to maintain the critical liver functions during the infection. Any number less than 20\% is lethal for the sufferer. Thus, for further study we pay close attention to this optimum  value of  uninfected hepatocytes.

\section{How does the proliferation rate influence the infection?}
Both healthy and unhealthy hepatocytes have the ability to undergo proliferation in order to compensate the deficiency of the liver cells. In this study, we mainly  focus to explore the roles of proliferation of liver cells during HBV infection. The understanding of how proliferation affects chronic infections is relatively scarce compared to the extensive knowledge available for acute infections. However, this experiment seeks to investigate  the influence of proliferation on both acute as well as chronic infection. In order to achieve this purpose, we examine the impacts of proliferation from two distinct perspectives. In the first case, it is  assumed that both types of hepatocytes proliferate with same rate ($r=\rho$), and secondly that the rate differs ($r \neq \rho$).  In this context, four distinct cases are arrived as follows:  
\begin{enumerate}[(i)]
	\item[\textbf{Case-1:}]  Same proliferation rates of uninfected and infected hepatocytes in  acute infection.
	\item[\textbf{Case-2:}]  Different proliferation rates of uninfected and infected hepatocytes in  acute infection.
	\item[\textbf{Case-3:}] Same proliferation rates of uninfected and infected hepatocytes in  chronic infection.
	\item[\textbf{Case-4:}]  Different proliferation rates of uninfected and infected hepatocytes in  chronic infection.
\end{enumerate}
\begin{figure}[h!]
	\centering
	\includegraphics[height=9cm,width=15cm]{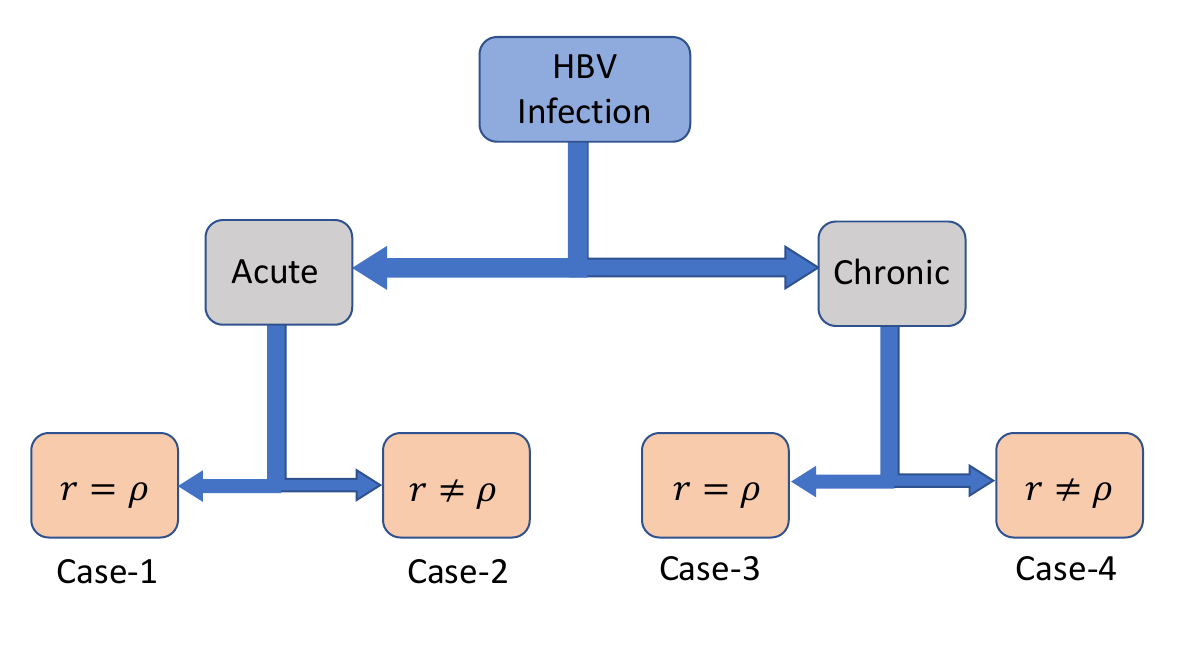}
	\caption{Schematic representation of four cases.}
	\label{}
\end{figure}
\subsection{Effects of same proliferation rates on  acute  and chronic infection $(r=\rho)$}
In this case, it is considered that both types of hepatocytes proliferate with same rate $(r)$. Accordingly,  equations \eqref{eq_susceptible hepatocytes} and \eqref{eq_infected hepatocytes} reduce to the following forms: 
\begin{equation} \label{eq_susceptible hepatocytes r=rho}
	\left.
	\begin{split}
		\dfrac{dX}{dt}&=r (X+2Y)\left(1-\dfrac{X+Y+\theta T}{T}\right)-k X V-\mu X\\
			\dfrac{dY}{dt}&=k X V-r Y \left(1-\dfrac{X+Y+\theta T}{T}\right)-\delta  Y
	\end{split}
\right\},
\end{equation}
where  equations (\ref{eq_capsids} and \ref{eq_virus}) remain same. 
\subsubsection{Steady states}
The steady states of the system (equations \eqref{eq_capsids},\eqref{eq_virus} and \eqref{eq_susceptible hepatocytes r=rho})  are obtained by solving  the system of  equations given by

\begin{align} 
		r (X+2Y)\left(1-\dfrac{X+Y+\theta T}{T}\right)-k X V-\mu X=0, \label{equilibrium_uninfected}\\
		k X V-r Y \left(1-\dfrac{X+Y+\theta T}{T}\right)-\delta  Y=0,\label{equilibrium_infected}\\
		a Y+\gamma(1-\alpha)D-\alpha\beta D-\delta D=0, \label{equilibrium_capsids}\\
		\alpha \beta D-cV=0. \label{equilibrium_virus}
\end{align}
\begin{enumerate}[(i)]
	\item \textbf{Trivial  or liver failure steady-state:} Clearly, $E_0=(0,0,0,0)$ is a solution of equations \eqref{equilibrium_uninfected}-\eqref{equilibrium_virus}. 
	\item \textbf{Disease-free steady-state:} Substituting $Y=0,~D=0,~$ and $V=0$ in equation \eqref{equilibrium_infected}-\eqref{equilibrium_virus}, we obtain the non-zero value of $X$ as $\dfrac{T(r-r\theta-\mu)}{r}$. Therefore, $E_f=\left( \dfrac{T(r-r\theta-\mu)}{r},0, 0, 0\right)$ is another steady state. In the context of biology, this steady state will exist  if and only if $(r-r\theta-\mu)> 0$.
	\item \textbf{Fully infection steady-state:} Substituting $X=0$ in equation \eqref{equilibrium_uninfected}, one can get $Y=(T-\theta T)$, but this value of $Y$ doesn't satisfy the equation \eqref{equilibrium_infected}. So, there does not exist any fully infected steady state.
	\item \textbf{Endemic steady-state:} The system has two endemic steady-states. The individual expressions of these two equilibria are not  given here due to the complexity.
\end{enumerate}
 The basic reproduction number of the model is calculated using next-generation approach \cite{2015_martcheva_introduction} and given by  
 $$R_0=\dfrac{a \alpha  \beta  k T (r-r \theta -\mu)}{rcR_s(\delta+\mu)}.$$
\subsubsection{Stability of equilibrium points}
\begin{figure}[h!]
	\centering
	\includegraphics[height=5cm,width=18cm]{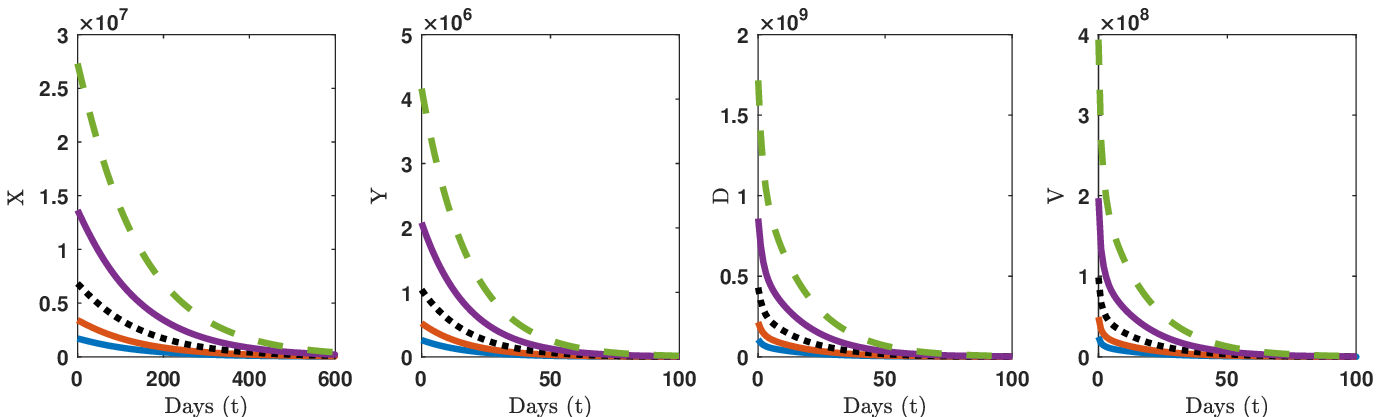}
	\caption{Convergence of solution to the liver failure equilibrium point for five different initial conditions.}
	\label{Fig: liver_failure}
\end{figure}

\begin{itemize}
	\item \textbf{Liver failure equilibrium point}
\end{itemize}
Acute liver failure (ALF) also known as fulminant hepatic failure, is the loss of liver function that happens quickly, usually in a matter of days or weeks, and in people who do not already have liver disease. Although ALF is quite uncommon. Most frequently, the main cause of ALF are  the use of drugs  or the hepatitis virus infection. In literature, it is seen that  chronic liver failure (CLF), which develops itself more slowly, is more common than acute liver failure. In order to study the stability of this steady-state, we first calculate the Jacobian matrix at $E_0$, and it is given by
 \begin{equation}
	J_{E_0}=
	\begin{pmatrix}
		r (1-\theta )-\mu  & 2 r (1-\theta ) & 0 & 0 \\
		0 & -\delta -r (1-\theta ) & 0 & 0 \\
		0 & a & -\alpha  \beta +(1-\alpha ) \gamma -\delta  & 0 \\
		0 & 0 & \alpha  \beta  & -c \\
	\end{pmatrix}.
 \end{equation}
 The corresponding eigenvalues are 
 $\{-c,-\alpha  \beta +(1-\alpha ) \gamma -\delta ,-\delta -(1-\theta ) r,(1-\theta ) r-\mu \}$. All eigenvalues except $(r(1-\theta )-\mu)$ are negative. The eigenvalue $(r(1-\theta )-\mu)$ will be negative if $(1-\theta ) r<\mu$. Hence, liver failure equilibrium  will be locally asymptotically stable if $2r<5\mu$. In Figure \ref{Fig: liver_failure}, the trajectories of both uninfected and infected hepatocytes, capsids and viruses are presented  considering five different initial conditions. In this case, it is seen that when the parameters $r$ and $\mu$ follow the inequality $(r(1-\theta )-\mu)<0$, the solution of the system converges to the liver failure equilibrium point.
 
\begin{remark}
In biological terminology, the inequality $2r<5\mu$ provides the following  crucial information: if the proliferation rate and the death rate of uninfected hepatocytes meet this inequality, the health condition of the patient will deteriorate day by day, and the liver will subsequently become damage permanently.
\end{remark}  

\begin{remark}
	In general, for all humans, the death or degradation rate of uninfected hepatocytes is almost constant. Humans have no control over it. Proliferation rates differ between individuals. In this circumstance, the only option to prevent liver failure is to keep the proliferation rate above a certain threshold value.  
\end{remark}
 
\begin{figure}[h!]
	\centering
	\includegraphics[height=4.5cm,width=15cm]{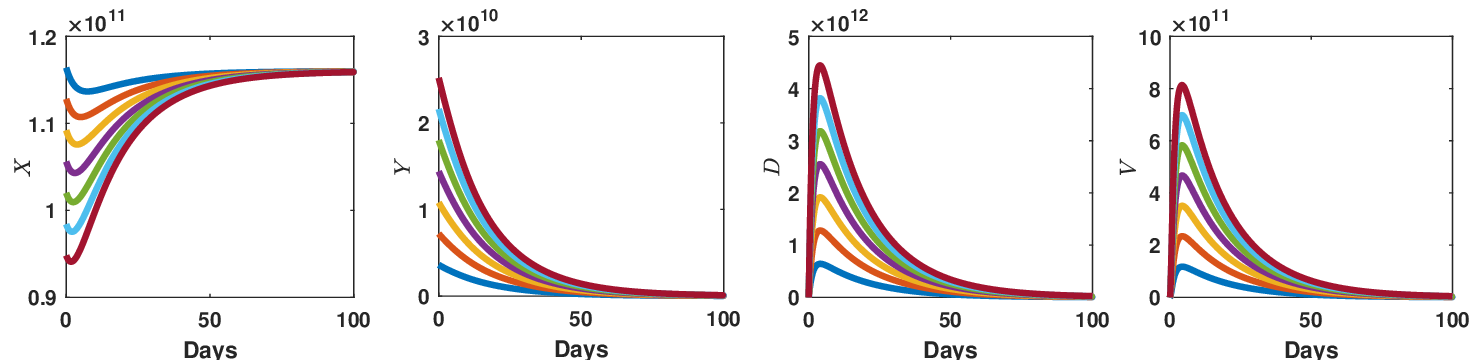}
	\caption{Times series of concentration of four variables $(X,~Y,~D,~V)$  for eight distinct initial conditions.}
	\label{fig:Equal_proliferation_rate_disease_free}
\end{figure}

\begin{itemize}
	\item \textbf{Disease-free steady-state}
\end{itemize}
The Jacobian matrix at $E_f$ is given by\\
\begin{equation}
J_{E_f}=
\begin{pmatrix} \label{matrix_diseasefree}
	r (\theta -1)+\mu  & r (\theta -1)+3 \mu  & 0 & -\dfrac{k T( r- \theta  r- \mu )}{r} \\
	0 & -(\delta+\mu)  & 0 & \dfrac{kT ( r- \theta  r- \mu )}{r} \\
	0 & a & -\alpha  \beta +(1-\alpha ) \gamma -\delta  & 0 \\
	0 & 0 & \alpha  \beta  & -c \\
\end{pmatrix}
\end{equation}
\noindent
Clearly, $r (\theta -1)+\mu$ is an eigenvalue of the matrix \eqref{matrix_diseasefree}, and it will be negative if $\mu<r(1-\theta)$. Other three eigenvalues are the eigenvalues of the matrix\\
\begin{equation}
\begin{pmatrix}\label{3*3 matrix}
	-(\delta +\mu)  & 0 & \dfrac{k T(r-\theta  r-\mu )}{r} \\
	a & -\alpha  \beta +(1-\alpha ) \gamma -\delta  & 0 \\
	0 & \alpha  \beta  & -c \\
\end{pmatrix}
\end{equation}
 Let the characteristic equation of this matrix \eqref{3*3 matrix} be
 \begin{align*}
 	\xi^3+\mathcal{A}_1 \xi^2+\mathcal{A}_2 \xi+\mathcal{A}_3=0, \\
 \end{align*}
where
\begin{align*}
\mathcal{A}_1&= \alpha  \beta +(\alpha -1) \gamma +c+2 \delta +\mu=R_s+\delta+\mu>0~\text{when}~R_s>0,\\
\mathcal{A}_2&= R_s(\delta+\mu+c)+c\delta+c\mu>0~\text{when}~R_s>0,\\
\mathcal{A}_3&= a \alpha  \beta  \theta  T k-a \alpha  \beta  T k+\frac{a \alpha  \beta  T k \mu }{r}+c(\delta+\mu)R_s>0, ~\text{when}~R_s>0,~ R_0<1,\\
\mathcal{A}_1 \mathcal{A}_2-\mathcal{A}_3&=\dfrac{r R_s(\delta +\mu )  (R_s+ \delta +\mu )-a \alpha  \beta  T k (\mu +(\theta -1) r)}{r}\\&+c^2 (R_s+ \delta +\mu )+c (R_s+ \delta +\mu )^2>0~\text{when}~R_s>0,~ R_0<1.
\end{align*}	
So, using \textit{Routh Hurwitz} criteria \cite{2015_martcheva_introduction}, we have the disease-free equilibrium is locally stable when the following conditions are satisfied: 
 \begin{multicols}{3}
	\begin{enumerate}[i.]
		\item $\mu<r(1-\theta).$
		\item $R_s>0.$
		\item $R_0<1$.		
	\end{enumerate}
\end{multicols}
\noindent
In addition, the system is numerically solved under eight different initial conditions, and in each instance, the system stabilizes to the disease-free steady-state as shown in Figure \ref{fig:Equal_proliferation_rate_disease_free}.
\noindent
\begin{remark}
	The stability criteria ($(1-\theta)r-\mu<0$ for liver failure, $(1-\theta)r-\mu>0$ for disease-free steady-states) indicate that the proliferation of hepatocytes has a significant influence on the dynamics of infection.
\end{remark}
\noindent	
The stability of the other two equilibrium points can not be investigated analytically due to the high complexity of the model. 
\subsubsection{Effects of proliferation rate }
The proliferation of infected hepatocytes can vary between acute and chronic infections.
One of the prime objective of this study is to determine how hepatocyte proliferation contributes to HBV infection. To study the effects of proliferation rate, we  set five different values of proliferation rate $(r)$ within realistic range, as mentioned in Table \ref{Table-Parameter values}. In addition, the possible relationships between initial situation of the patients and proliferation  are also explored for both acute and chronic infection.

\subsubsection*{Acute infection}
\begin{figure}[h!]
	\centering
	\includegraphics[width=16cm,height=7cm]{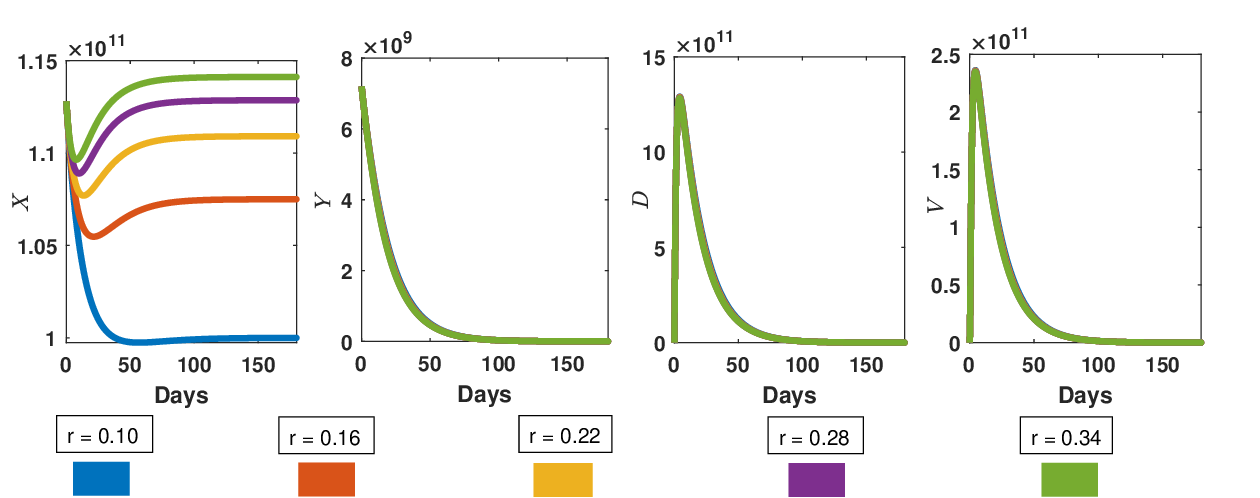}
	\caption{Effects of proliferation rate in acute infection.}
	\label{Proliferation_rate_acute}
\end{figure}

Figure \ref{Proliferation_rate_acute} illustrates how the infection is influenced by the simultaneous proliferation of infected and uninfected hepatocytes. In this case, the parameter values are chosen in such a way that the value of $R_0$ remains below unity. When proliferation rate $(r)$ equal to 0.1 (signifying negligible proliferation), maximum liver cell becomes infected within few months. For this value of  proliferation rate, the number of uninfected cells falls below the minimum  required level for the liver to function smoothly, as mentioned in Section \ref{smoothness of the liver}. The number of uninfected hepatocytes increases at a fixed time as the proliferation rate rises.  Therefore, examining the profile of  uninfected hepatocyte, it can be concluded that the  proliferation reduces infection in acute case.  



\subsubsection*{Chronic infection}
\begin{figure}[h!]
 	\centering
 	\includegraphics[width=16cm,height=7cm]{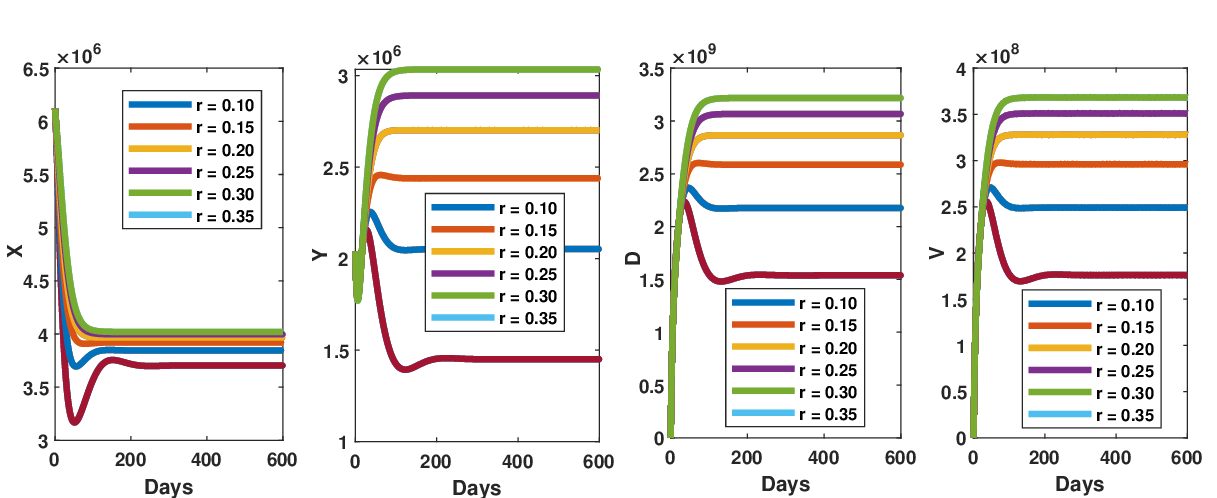}
 	\caption{Effects of proliferation rate on chronic infection.}
 	\label{Proliferation_rate_chronic}
 \end{figure}

In the case of chronic infection, we have examined the impacts of the proliferation rate and  explored the relationship between proliferation and the initial phage of the sufferers. The proposed model is solved for different values of proliferation rate $(r)$ and the solutions are plotted  in Figure \ref{Proliferation_rate_chronic}.  As observed in acute infection, the growth of proliferation rate leads to negligible change in the concentration  of infected hepatocytes, capsids and virus compartments  but in this case the change is substantial and considerable. Infected hepatocytes exhibit a more pronounced change in their profile compared to uninfected counterparts.
Since the stability levels of infected hepatocytes, capsids, and viruses rise significantly with the increase of proliferation as  depicted in the Figure \ref{Proliferation_rate_chronic}, it is conspicuous that proliferation serves as a positive factor in the persistence of the infection.
%
%

\begin{center}
 \begin{figure}[h!]
\includegraphics[width=16cm,height=10cm]{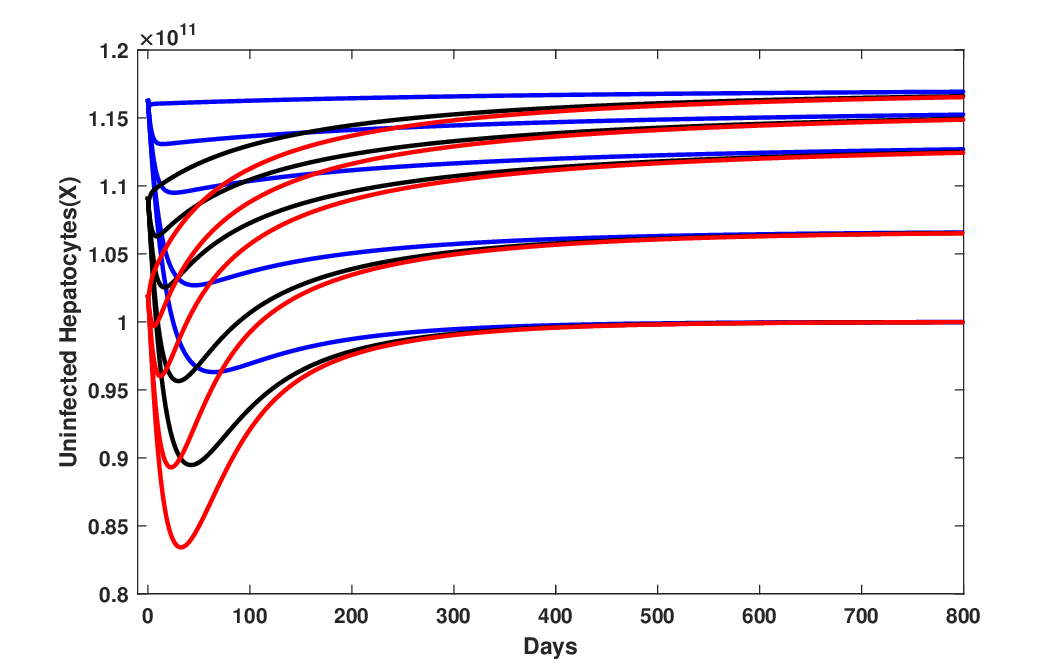}
\caption{The correlation between the  proliferation and initial condition of the patients.}
\label{Proliferation_rate_chronic_initial}
\end{figure}
 	\end{center}

In order to elucidate the association between initial condition of the patients and the proliferation, five different initial conditions $(ic_1,~ic_2,~ic_3,~ic_4,~ic_5)$ are considered, and the simulation results are presented in Figure \ref{Proliferation_rate_chronic_initial}. For all initial states, proliferation seems to play equal roles. The long-term behavior of the infection is solely governed by mitosis. We don't observe any notable role of  the initial condition of the patients in the chronicity of the infection. The conclusions, we draw are grounded on the results of a finite number of simulations, taking into account the smooth functional criteria of the liver outlined in the Section \ref{smoothness of the liver}. To better understand the roles of patients' initial conditions globally, further research  is imperative.
%

\subsection{Effects of different proliferation rates on the infection $(r\neq\rho)$}
In this section, it is considered that the rate at which infected hepatocytes proliferate is lower than the rate of proliferation observed in healthy hepatocytes, \textit{i.e.}, $\rho<r$ \cite{2021_hews_global}. As a result, the reaction equations corresponding to uninfected and infected hepatocytes transform into the following equations:
\begin{equation} \label{eq_susceptible hepatocytes r not equal to rho}
	\left.
	\begin{split}	
		\dfrac{dX}{dt}&=r X \left(1-\dfrac{X+Y+\theta T}{T}\right)+2 \rho Y \left(1-\dfrac{ X+Y+\theta T}{T}\right)-k X V-\mu X,\\
		\dfrac{dY}{dt}&=k X V-\rho Y \left(1-\dfrac{X+Y+\theta T}{T}\right)-\delta  Y.
	\end{split}
\right\},
\end{equation}
\noindent
In Section \ref{Sec: Model formulation and analysis}, we have already discussed that  this system of equations (equations \eqref{eq_susceptible hepatocytes r not equal to rho}, \eqref{eq_capsids} and \eqref{eq_virus}) possesses a distinct, bounded, and non-negative solution initiating from a non-negative initial condition. Our main cynosure of this section lies on understanding the infection dynamics influenced by the different proliferation rates of uninfected and infected cells. We rigorously explore how different proliferation rates $(r,~\rho)$  affect the dynamics of infection in the present of standard incidence function and capsid recycling. In the following section, a thorough study is carried out on the proliferation of both types of hepatocytes.
\subsubsection{Acute infection} \label{Acute infection}
\begin{figure}[h!]
	\centering
	\includegraphics[width=15cm, height=10cm]{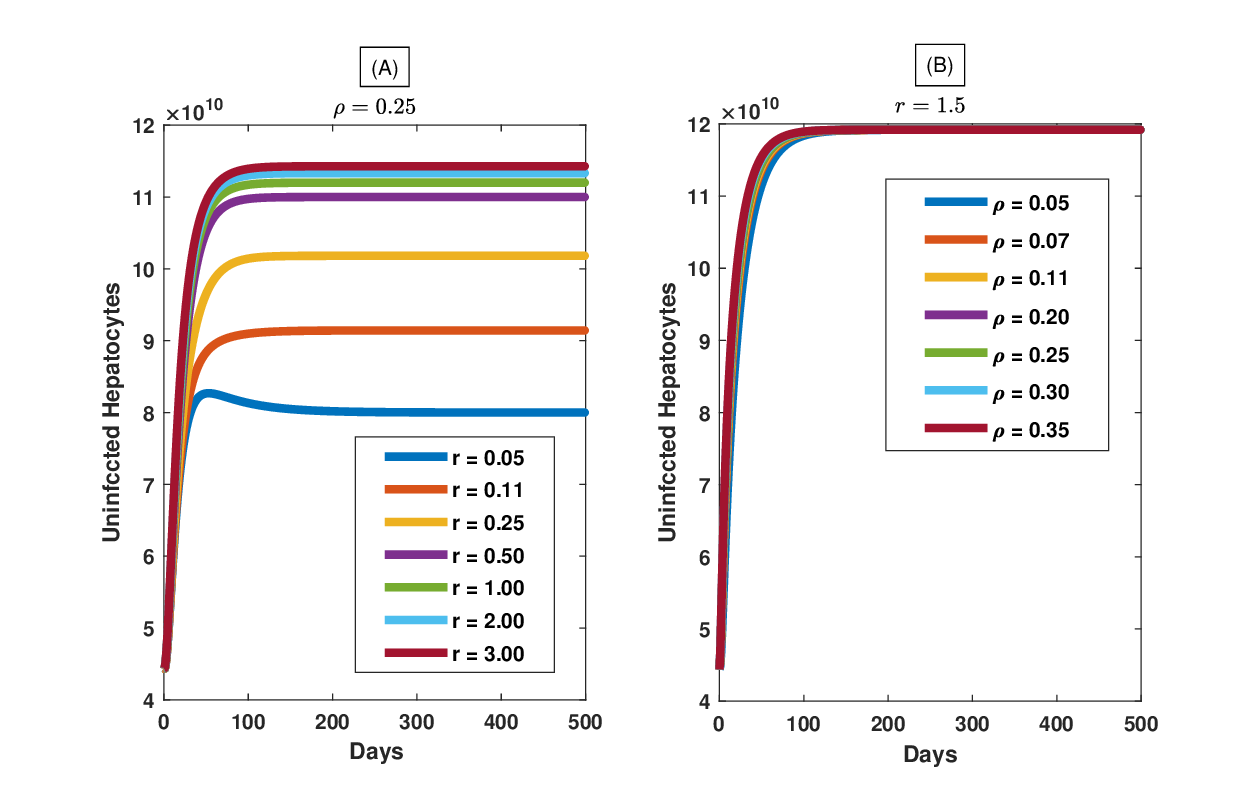}
	\caption{Effects of different proliferation rates on acute infection. (A): The impacts of $r$. (B): The impacts of $\rho$.}
   \label{Proliferation_rate_acute_unequal}
\end{figure}

In this case, we have conducted two experiments:
\begin{itemize}
	\item \textbf{Experiment-1:}  Proliferation rate of uninfected hepatocytes $(r)$ varies within a reasonable range $(0.001-3.4)$ and  proliferation rate of infected hepatocytes is kept fixed at $\rho=0.25$ (average value, according to Table \ref{Table-Parameter values}).  
	\item \textbf{Experiment-2:}  Proliferation rate of infected hepatocytes $(\rho)$ varies within a reasonable range $(0.001-0.35)$ and  proliferation rate of uninfected hepatocytes is kept fixed at $r=1.7$ (average value, according to Table \ref{Table-Parameter values}). 
\end{itemize}
Figures \ref{Proliferation_rate_acute_unequal}(A) and \ref{Proliferation_rate_acute_unequal}(B) display the simulation results for Experiment-1 and Experiment-2, respectively. Since the severity of the infection is generally determined by the number of uninfected hepatocytes, we only focus on the profile of uninfected cells. In the Experiment-1, it is observed that when uninfected hepatocytes undergo rapid proliferation, there is a significance increase in the concentration  of healthy hepatocytes. Consequently, the disease gradually diminishes over time.

On the other hand, in case of  Experiment-2, the trend is clear. If only  infected hepatocytes are allowed to proliferate, the number of uninfected hepatocytes   convergences swiftly  to the infection-free steady-state, resulting in a speedy recovery from the infection. 
\begin{remark}
	The changes in the profile of healthy cells due to variations in $\rho$ are negligible compared to the changes observed in the case of variations in $r$. However, the rapid proliferation of both cells reduce the severity of the infection.
\end{remark}
\subsubsection{Chronic infection} 
\begin{figure}[h!]
	\centering
	\includegraphics[width=15cm, height=12cm]{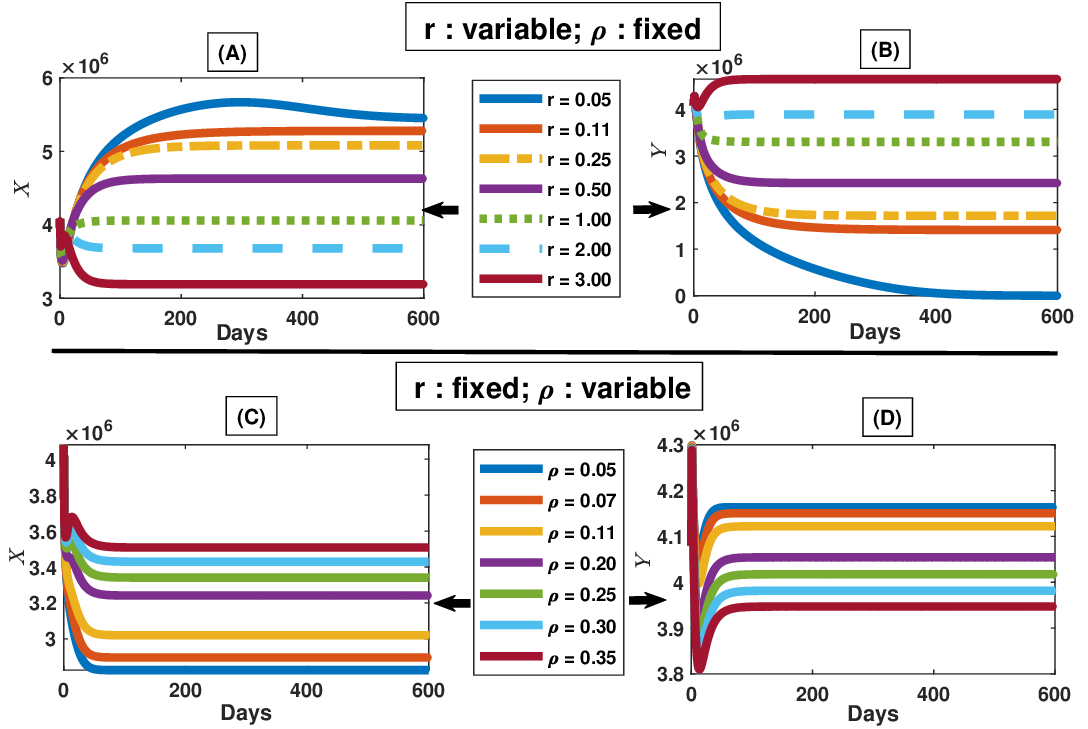}
	\caption{Effects of different proliferation rate $(r\neq \rho)$ in chronic infection.}
	\label{Proliferation_rate_chronic_unequal}
\end{figure}

The  understanding of liver cell proliferation during chronic infection is still incomplete and  inadequate. It also remains unclear whether the roles of proliferation in chronic infections  resemble those observed in acute infections or not. In order to explore the biologically plausible dynamics in chronic infection,  two more experiments are carried out similarly to the case discussed above.
\begin{itemize}
	\item \textbf{Experiment-3 ($r$ : free variable, $\rho$ : fixed) :} The influence of proliferation of uninfected hepatocytes  are discussed in the context of long-term infection. In Figures \ref{Proliferation_rate_chronic_unequal}(A) and \ref{Proliferation_rate_chronic_unequal}(B), the dynamic temporal behavior of uninfected and infected hepatocytes are demonstrated. Compare to the acute infection, chronic infections manifest an opposite trend on both compartments.  In this circumstance, the infection becomes more severe due to rapid proliferation.
	  Based on the Figures \ref{Proliferation_rate_chronic_unequal}(A) and  \ref{Proliferation_rate_chronic_unequal}(B) and rigorous analysis of the proposed model, 
the probable reason behind this opposite trend in the profiles of liver cells can be clarified by considering the fact that the rapid proliferation of uninfected hepatocytes results quick increase in the population of healthy hepatocytes. Consequently, this leads to the frequent interaction between viruses and healthy cells which intensifies infection. 
	\item \textbf{Experiment-4: ($r$: fixed, $\rho$: free variable)} In this case, we study how the proliferation of infected hepatocytes shapes various outcomes.  The correlation between infection and the proliferation of infected hepatocytes during the infection is highlighted based on numerical results. The variations in the outcomes of uninfected and infected liver cells are visualized in Figures \ref{Proliferation_rate_chronic_unequal}(C) and \ref{Proliferation_rate_chronic_unequal}(D). The increase in the rate of infected cells proliferation leads to an increase in the number of uninfected cells and a decrease in the number of infected cells. Therefore, the persistence of chronic infection is  closely associated with the cellular proliferation  of infected cells resulting in two uninfected cells.
\end{itemize}
For our easily understanding, all the  results obtained up-to this section are put together in Table \ref{all results}. With the increase of proliferation rates, if the number of hepatocytes increases, we enter ``Increase" in the corresponding cell of the Table \ref{all results}. In case of  decrease in hepatocytes, we put ``Decrease".
\begin{center}
	\begin{table} 
		\begin{tabular}{|ll|c|cc|}
			\hline
			\multicolumn{2}{|l|}{\multirow{2}{*}{}}    & \multirow{2}{*}{$r=$ $\rho$ (increase)} & \multicolumn{2}{c|}{$r$ $\neq$ $\rho$}    \\ \cline{4-5} 
			\multicolumn{2}{|l|}{}                     &                   & \multicolumn{1}{l|}{$r$ increase, $\rho$  fixed} & $\rho$ increase, $r$  fixed \\ \hline
			
			\multicolumn{1}{|l|}{\multirow{2}{*}{Acute}} & Uninfected hepatocytes & Increase                  & \multicolumn{1}{c|}{Increase} & Increase  \\ \cline{2-5} 
			\multicolumn{1}{|l|}{}                  &Infected hepatocytes  &  Decrease                 &  \multicolumn{1}{c|}{Increase} &  Decrease \\ \hline
			
			\multicolumn{1}{|l|}{\multirow{2}{*}{Chronic}} &  Uninfected hepatocytes & Increase                  & \multicolumn{1}{c|}{Decrease} & Increase  \\ \cline{2-5} 
			\multicolumn{1}{|l|}{}                  &Infected hepatocytes  & Increase                  & \multicolumn{1}{c|}{Increase} & Decrease \\ \hline
		\end{tabular}
	\caption{The correlation between proliferation of both types of hepatocytes (uninfected and infected) and  acute as well as chronic infections.}
	\label{all results}
	\end{table}
\end{center}

\section{Modified model: Proliferation is not a continuous biological process}
\begin{figure}[h!]
	\centering
	\includegraphics[width=15cm, height=7cm]{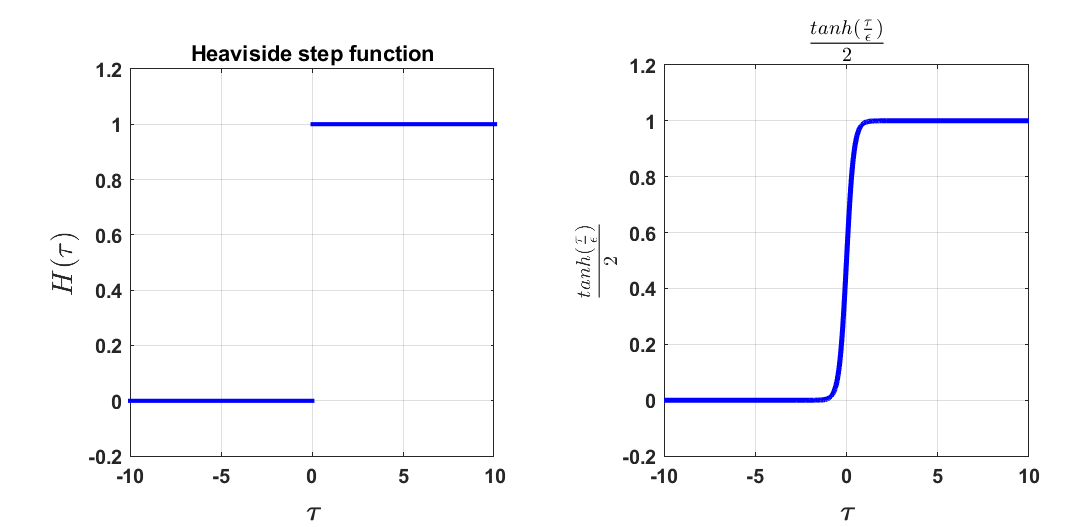}
	\caption{Heaviside step function and its approximation function.}
	\label{heaviside_tanh}
\end{figure}

  In the proposed model (equations \eqref{eq_susceptible hepatocytes}, \eqref{eq_infected hepatocytes}, \eqref{eq_capsids} and \eqref{eq_virus}), it was assumed that proliferation always occurs  \textit{i.e.} proliferation is a continuous biological process. 
  In the literature it is seen that  during liver regeneration after partial resection or injury, the remaining hepatocytes rapidly multiply to restore liver function. In response to liver damage or diseases like hepatitis or cirrhosis, compensatory hepatocyte proliferation occurs to replace damaged cells. According to a study of Miyaoka and Miyajima \cite{2013_miyaoka_divide}, cellular proliferation is stimulated solely when the total number  liver cell population decreases to less than 70\% of its initial volume. In order to include this phenomenon in the proposed model,  the standard proliferation terms  $\left(r X+2\rho Y\right)\left(1-\dfrac{X+Y+\theta T}{T}\right)$ and $\rho Y\left(1-\dfrac{X+Y+\theta T}{T}\right)$   are replaced by   $$\left(r X+2\rho Y\right) \left(1-\dfrac{X+Y+\theta T}{T}\right)H\left(0.7-\dfrac{X+Y+\theta T}{T}\right)$$
   and 
   $$\rho Y\left(1-\dfrac{X+Y+\theta T}{T}\right)H\left(0.7-\dfrac{X+Y+\theta T}{T}\right),$$
    respectively, where $H(\tau)$ denotes Heaviside step function defined by 
 $$H(\tau)=
 \begin{cases}
       0,~ \tau <0,\\
       1,~ \tau \geq 0.
 \end{cases}
 $$ 
 The associated modified governing equations for uninfected and infected classes  are given by the following equations:
 \begin{equation} \label{modified model}\
 	\left.
 	\begin{split}
 		\dfrac{dX}{dt}&=\left(r X+2\rho Y\right)\left(1-\dfrac{X+Y+\theta T}{T}\right)H\left(0.7-\dfrac{X+Y+\theta T}{T}\right)-k X V-\mu X,\\
 		\dfrac{dY}{dt}&=k X V-\rho Y  \left(1-\dfrac{X+Y+\theta T}{T}\right)H\left(0.7-\dfrac{X+Y+\theta T}{T}\right)-\delta  Y. 		
 	\end{split}
 \right\}
 \end{equation}
Since, $H(\tau)$ is a discontinuous function at $\tau=0$, its applicability in continuous systems is restricted as  it is unable to capture accurately an uninterrupted process. So, we approximate $H(\tau)$ by a smooth function as
$$H(\tau)\approx\dfrac{1}{2}\left(1+\tanh\left(\frac{\tau}{\epsilon}\right)\right),$$ where $\epsilon$ is very small in magnitude.  The function $\left(1+\tanh\left(\frac{\tau}{\epsilon}\right)\right)$  provides a
smooth, monotonic description of proliferation. The graphical representation of these two functions are shown in Figure \ref{heaviside_tanh}. The capsids and virus classes remain unchanged; no modifications are made in this context. 
\subsection{Regular proliferation vs irregular proliferation}
\begin{figure}[h]
	\centering
	\includegraphics[width=15cm, height=10cm]{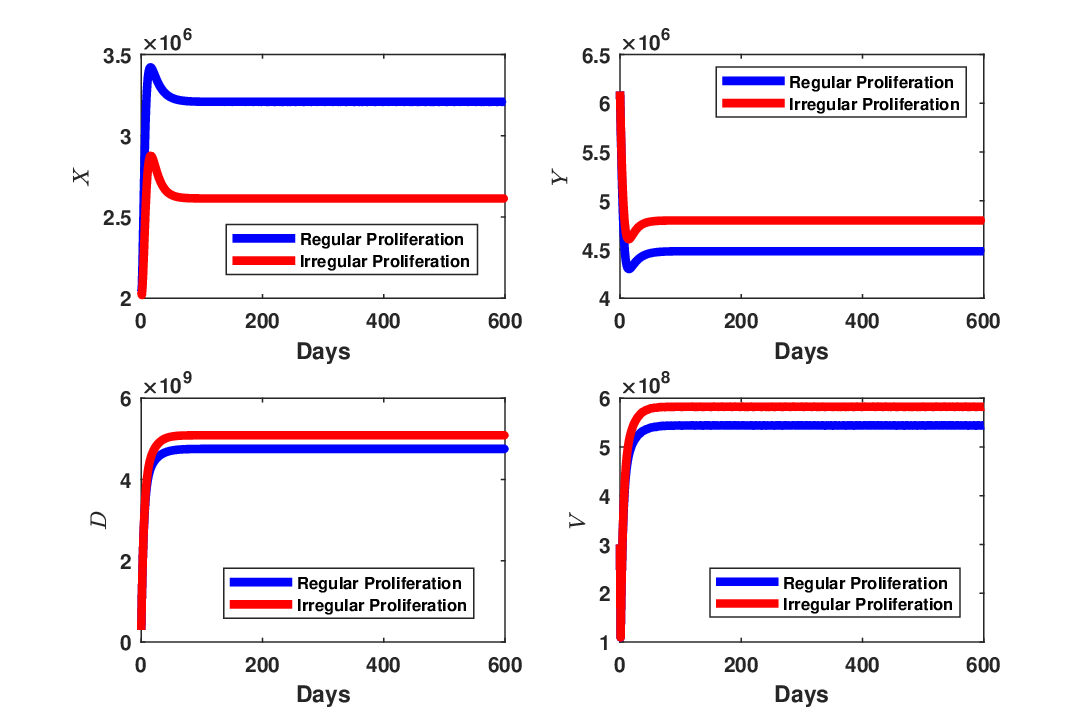}
	\caption{Regular proliferation vs irregular proliferation.}
	\label{proliferation70}
\end{figure}

\begin{justify}

For the convenience of our discussion, the following things are defined as: 
\begin{itemize}
	\item \textbf{Regular or standard proliferation:} Both the terms $r X\left(1-\dfrac{X+Y+\theta T}{T}\right)$ and $\rho Y\left(1-\dfrac{X+Y+\theta T}{T}\right)$ are defined as regular or standard  proliferation due to the genesis of these terms.
	\item \textbf{Irregular proliferation:}  The expressions  $r X\left(1-\dfrac{X+Y+\theta T}{T}\right)H\left(0.7-\dfrac{X+Y+\theta T}{T}\right)$ and $\rho X\left(1-\dfrac{X+Y+\theta T}{T}\right)H\left(0.7-\dfrac{X+Y+\theta T}{T}\right)$ used in the equation \eqref{modified model} are termed as irregular proliferation  because it occurs only when the number of liver cell falls below 70\% of the total number of liver cells.
\end{itemize}
We also define the system of equations (\eqref{eq_susceptible hepatocytes}, \eqref{eq_infected hepatocytes}, \eqref{eq_capsids} and \eqref{eq_virus}) as model-p (primary model) and system of equations (\eqref{modified model},  \eqref{eq_capsids} and \eqref{eq_virus}) as model-m (modified model). Both models (model-p and model-s) are solved numerically due to the presence of high non-linearity, and the corresponding solutions are demonstrated in Figure \ref{proliferation70}. As a result, it is observed that despite the consideration of same parameters values and a same fixed initial condition, there are substantial variations in the solutions. It is also noticed that in case of irregular proliferation, the infection becomes more severe.  
Although much information are unexplored in the context of irregular proliferation, these findings hold the promise of enriching our prior knowledge. However, a more comprehensive and detailed study is imperative in this area.
\end{justify}
\section{Conclusions}
In this paper, the effects of proliferation of uninfected and infected hepatocytes on hepatitis B viral infection are studied by considering all possible cases.  The recycling of capsids is also incorporated in this model. Capsids recycling is one of the  distinctive features in the proposed model. We prove that the solution of the system starting from non-negative initial condition remains non-negative and bounded. The stability of the equilibrium points are established based on the value of basic reproduction number and proliferation rate. This study also addresses the consequences of both regular and irregular proliferation for the first time.

The results of our study reveal the following key findings.
\begin{enumerate}
	\item When both uninfected and infected hepatocytes proliferate with equal rate $(r = \rho)$ in acute infection, the number of uninfected hepatocytes increases as the proliferation rate rises. At the same time, the number of infected hepatocytes decreases to zero. This suggests that during acute infection, proliferation aids the individual in a rapid recovery from the infection. 
	
	 In case of chronic infection, the severity of the infection increase if the proliferation occur frequently.
    \item When uninfected and infected hepatocytes proliferate with different rates $(r\neq \rho)$, proliferation of uninfected hepatocytes cure the disease quickly where we do not find  any significant contribution of proliferation of infected hepatocytes in acute case.
    
    On the other hand, the experiments on chronic infection indicate that the infection is highly influenced by proliferation of both types of cells. The proliferation of uninfected increase the infection whereas proliferation of infected hepatocytes decrease the infection. 
    
	\item The difference between the solutions of regular and irregular proliferation is substantial and noteworthy. In order to determine which solution better approximates reality, additional research and investigation are needed.	
\end{enumerate}
The analysis of our study represents some testable biological findings. This research provides greater insights and holds more significance. It is expected that this model will be used in a variety of purposes such as, in the development of any novel therapeutics strategies, future studies,  clinical trials.
\subsection*{Availability of data and materials}
\noindent
Data sharing is not applicable to this article.
\subsection*{Competing interests}
\noindent
The authors declare that they have no competing interests.
\subsection*{Authors' contributions}
\noindent 
Both authors contribute equally.
\subsection*{Acknowledgments}
\noindent
First author would like to acknowledge the financial support obtained from CSIR (New Delhi) under
the CSIR-SRF Fellowship scheme (File No: 09/731(0171)/2019-EMR-I). The first author also thanks
the research facilities received from the Department of Mathematics, Indian Institute of Technology
Guwahati, India. 
\end{document}